\documentclass[twocolumn]{webofc}
\usepackage[varg]{txfonts}   % Web of Conferences font
\usepackage{cancel}
\begin{document}
\title{Behavior of the continuum coupling correlation energy in the vicinity of the particle emission threshold - Gamow shell model study}

\author{\firstname{Jose Pablo} \lastname{Linares Fernandez}\inst{1,2}\fnsep\thanks{\email{jose.linares386@gmail.fr}} \and
        \firstname{Nicolas} \lastname{Michel}\inst{3,4}\fnsep\thanks{\email{nicolas.lj.michel@outlook.fr}} \and
        \firstname{Marek} \lastname{P{\l}oszajczak}\inst{2}\fnsep\thanks{\email{marek.ploszajczak@ganil.fr}}
        % etc.
}

\institute{Department of Physics and Astronomy, Louisiana State University, Baton Rouge, LA 70803, USA 
 \and
 Grand Acc\'el\'erateur National d'Ions Lourds (GANIL), CEA/DSM - CNRS/IN2P3, BP 55027, F-14000 Caen, France
\and
           AS Key Laboratory of High Precision Nuclear Spectroscopy, Institute of Modern Physics, Chinese Academy of Sciences, Lanzhou 730000, China
 \and
 School of Nuclear Science and Technology, University of Chinese Academy of Sciences, Beijing 100049, China
           }

\abstract{%
The Gamow shell model provides the open quantum system formulation of nuclear shell model.  In the coupled-channel representation, Gamow shell model  provides the unified theory of nuclear structure and reactions which is well suited for the study of resonances and clusterization. In this work, we apply this approach to study the continuum-coupling correlation energy for selected near-threshold states of $^7$Li, $^7$Be
using a translationally invariant Hamiltonian with an effective finite-range two-body interaction. 
}
\maketitle
\section{Introduction}
\label{intro}
Properties of the radioactive nuclei are essential for understanding various astrophysical processes, in particular the nucleosynthesis. These nuclei are strongly affected by couplings to many-body continuum of scattering states and decay channels. Therefore, for the comprehensive study of radioactive nuclei it  is essential that bound states, resonances and scattering states are described within a single theoretical framework which unifies the description of nuclear structure and reactions. 
Such a framework is provided by  the Gamow shell model (GSM) \cite{michel_review_2009,michel_book_2021}, which offers the most general treatment of couplings between discrete and scattering states, as it makes use of Slater determinants defined in the Berggren ensemble of single-particle states~\cite{berggren_1968}. In the coupled-channel representation (GSM-CC)~\cite{jaganathen_2014,mercenne_2019} (see also Ref.~\cite{michel_book_2021}), the GSM describes asymptotics of the reaction channels and offers the unification of nuclear structure and reactions within the GSM theory. In the present study,  to solve the Schr{\"o}dinger equation, we will consider the binary mass partitions and the effective nucleon-nucleon interaction among valence nucleons outside the $^4$He core.  

One way to connect cross-sections to nuclear structure is via the spectroscopic factors (SFs). If nuclear structure and reactions are described in an open quantum system (OQS) framework, then the threshold effect in the cross-section is associated with an analogous effect in the SFs. Previous 
GSM studies \cite{michel2007continuum,michel2007threshold,michel2010isospin} have shown that the coupling to the continuum is indispensible to reproduce Wigner cusps in the structure calculations.

Another near-threshold effect is related to the build up of specific nucleon-nucleon correlation or clustering which share many properties with the nearby reaction channel. Long time ago, Ikeda \textit{et al.} \cite{ikeda1968systematic} have noticed that $\alpha$-cluster states exist close to the $\alpha$-particle emission threshold in many '$N\alpha$-nuclei'. In Refs. \cite{Okolowicz2012,Okolowicz2013}, it was conjectured that the near-threshold clustering is a general OQS phenomenon which might exist at any particle emission-threshold, even the one where the emitted particle is unbound, like the dineutron or diproton.  Actually, narrow near-threshold resonances are seen in atomic nuclei. 
In the shell model embedded in the continuum \cite{Okolowicz2003}, the recent version of the real-energy continuum shell model, studies of the near-threshold effects that may lead to clusterization, have been done using the  continuum-coupling correlation energy \cite{okolowicz2006continuum,okolowicz2007continuum}. In the GSM-CC framework, one can define a similar, albeit not the same, function which allows to investigate the role of different reaction channels in the structure of wave functions. 

The goal of this paper is to lay ground for the GSM-CC studies of the near-threshold phenomena by investigating the dependence of the continuum-coupling correlations on the energy difference from the particle emission threshold. We will present GSM-CC results in the case of $^7$Li and $^7$Be. The studies will be completed by a presentation of the near-threshold behavior of the squared amplitudes of various reaction channels in the  binary multi-mass partition coupled-channel framework of the GSM-CC.

\section{Theoretical framework}
\label{Formalism}
{
Below, we will briefly outline the GSM-CC formalism for the channels which are constructed with different mass partitions. Detailed discussion of the GSM-CC in a single mass partition case can be found in Refs.~\cite{jaganathen_2014,mercenne_2019,michel_book_2021}. The first application of the GSM-CC formalism with the multi-mass binary mass partitions have been presented in Ref.~\cite{Linares_2023}.
Here, we will mainly concentrate on the differences between single- and multi-mass partition case of the GSM-CC and apply for the description of selected states in $^7$Li and $^7$Be. 

We work in the cluster orbital shell model formalism (COSM) \cite{Suzuki1988}, where all space coordinates are defined with respect to the center-of-mass (c.m.)~of an inert core. For valence nucleons, one has $\mathbf{r} = \mathbf{r_{\rm lab}} - \mathbf{R_{\rm CM}^{\rm (core)}}$, where
$\mathbf{r_{\rm lab}}$ is the nucleon coordinate in the laboratory frame, $\mathbf{R_{\rm CM}^{\rm (core)}}$ is the core c.m.~coordinate in the laboratory frame, which then define $\mathbf{r}$ as the nucleon coordinate in the COSM frame. 
The fundamental advantage of COSM is that one deals with translationally invariant space coordinates, so that no spurious c.m. motion can occur~\cite{Suzuki1988,michel_book_2021}.
Additionally, the use of core coordinates greatly simplifies antisymmetry requirements, as Slater determinants built from one-body states in the COSM frame are both antisymmetric and without spuriosity~\cite{Suzuki1988,michel_book_2021}. 

The ${ A }$-body state of the system is decomposed into reaction channels defined as binary clusters:
\begin{equation}
  | { \Psi }_{ M }^{ J } > = \sum_{\rm  c } \int_{ 0 }^{ +\infty } |{ \left( {\rm c} , r \right) }_{ M }^{ J } > \frac{ { u }_{\rm c }^{JM} (r) }{ r } { r }^{ 2 } ~ dr \; ,
  \label{scat_A_body_compound}
\end{equation}
where the radial amplitude ${ {u}_{\rm c }^{JM}(r) }$, describing the relative motion between the two clusters in a channel ${\rm c }$, is the solution to be determined for a given total angular momentum ${J}$ and its projection ${M}$. 
The different channels in the sum of Eq.~(\ref{scat_A_body_compound}) are orthogonalized independently of their mass partition, involving different numbers of neutrons and protons. The integration variable ${ r }$ is the relative distance between the c.m.~of the cluster projectile and that of the inert core \cite{michel_book_2021}, and the binary-cluster channel states are defined as:
\begin{equation}
| \left( {\rm c} , r \right)> = \hat{ \mathcal{A}} [|\Psi_{\rm  T}^{J_{ \rm T }}; N_T, Z_T> \otimes  |r ~ L_{\rm CM} ~ J_{\text{\rm int}} ~ J_{\rm P}; n, z>]_{ M }^{J} \label{channel}
\end{equation} 
where the channel index ${\rm c}$ stands for different quantum numbers and mass partitions,  $N_T$ and $Z_T$ are the number of neutrons and protons of the target, and $n$ and $z$ are the number of neutrons and protons of a projectile, so that $N = N_T + n$ and $Z = Z_T + z$ are the total number of neutrons and protons in the combined system of a projectile and a target. ${\hat{ \mathcal{A}}}$ is the inter-cluster antisymmetrizer that acts among the nucleons pertaining to different clusters. 
The states $|\Psi_{\rm  T}^{J_{ \rm T }}> $ and $|r ~ L_{\rm CM} ~ J_{\text{\rm int}} ~ J_{\rm P}>$ are the target and projectile states in the channel $| \left( {\rm c} , r \right)>$ of Eq.~(\ref{channel}) with their associated total angular momentum ${ { J }_{ \rm T } }$ and ${ { J }_{ \rm P } }$, respectively.
The angular momentum couplings read $\mathbf{ J}_{\rm P } = \mathbf{ J}_{ \rm int } + \mathbf{L}_{\rm CM}$ and  ${ \mathbf{ J}_{\rm A} = \mathbf{J}_{\rm P} + \mathbf{ J}_{\rm T} } $. Quantum numbers of many-body projectiles are customarily denoted by $^{2J_{\rm int}+1}(L_{CM})_{J_{\rm P}}$ in numerical applications. These angular quantum numbers will also be denoted by $\ell j$ when dealing with one-nucleon systems.

The Schr{\"o}dinger equation $H |\Psi_{M}^{J}> = E |\Psi_{M}^{J}>$ in the channel representation of the GSM takes the form of coupled-channel equations:
\begin{equation}
  \sum_{\rm c}\int_{0}^{\infty}  \!\!\! r^{ 2 } \left( H_{\rm cc' } (r , r') - E N_{\rm cc' } (r , r') \right) \frac{ { u }_{\rm c } (r) }{ r } = 0	\ ,
  \label{cc_cluster_eq}
\end{equation}
where ${ E }$ stands for the scattering energy of the ${ A }$-body system. To simplify reading, we have dropped the total angular momentum labels ${ J }$ and ${ M }$, but one should keep in mind that Eq.~(\ref{cc_cluster_eq}) is solved for fixed values of ${J}$ and ${ M }$. The kernels in Eq.~(\ref{cc_cluster_eq}) are defined as:
\begin{align}
  & H_{\rm cc' } (r,r') = <({\rm c},r)| \hat{ H } |({\rm c'},r')> \label{h_cc_compound} \\
  & N_{\rm cc' } (r,r') = <({\rm c},r) | ({\rm c'},r') > \label{n_cc_compound}
\end{align}

As in our model the nucleons of the target and those of the projectile interact via a short-range interaction, it is convenient to express the Hamiltonian $\hat{ H }$ as: 
%$\hat{ H } = \hat{ H }_{ \rm T } + \hat{ H }_{ \rm P } + \hat{ H }_{ \rm TP}$~,
\begin{equation}
  \hat{ H } = \hat{ H }_{ \rm T } + \hat{ H }_{ \rm P } + \hat{ H }_{ \rm TP }	 \ ,	
  \label{new_hamiltonian}
\end{equation}
where ${ \hat{ H }_{ \rm T } }$ and ${ \hat{ H }_{ \rm P } }$ are the Hamiltonians of the target and projectile, respectively.
${ { \hat{ H } }_{ \rm T } }$ is the intrinsic Hamiltonian of the target, and its eigenvectors are ${ | { \Psi }_{ \rm T }^{ { J }_{ \rm T } } } >$ with eigenvalues ${ { E }_{ \rm T }^{ { J }_{ \rm T } } }$. ${ \hat{ H } }$ is considered to be the standard GSM Hamiltonian.
The projectile Hamiltonian ${ { \hat{ H } }_{ \rm P } }$ can be decomposed as follows: ${ { \hat{ H } }_{ \rm P } = { \hat{ H } }_{ \text{\rm int} } + { \hat{ H } }_{ \text{CM} } }$.
${ { \hat{ H } }_{\rm  int } }$ describes the intrinsic properties of the projectile and $|J_{\rm int}>$ is its eigenvector with an eigenvalue ${ { E }_{ \rm int }^{ { J }_{ \rm int } } }$.
${ { \hat{ H } }_{ \rm CM } }$ describes the c.m. of the projectile which in a single channel $c$ is defined  as:
 \begin{equation}
   { \hat{ H } }_{\rm CM } = \frac{ { \hbar }^{ 2 }}{ 2 \Tilde{m}_{\rm P}} \left( -\frac{ { d }^{ 2 }}{ d r^{ 2 }} + \frac{L(L + 1)}{r^2} \right)  + { U_{\rm CM}^{ L }} (r) \ ,
  \label{HCM_definition}
\end{equation} 
where $L=L_{\rm CM}$ is the c.m.~orbital angular momentum, ${ {\Tilde{m} }_{ \rm P } }$ is the reduced mass of the projectile. ${ { U }_{\rm CM }^{ L }(r) }$ in this expression is the basis-generating Woods-Saxon (WS) potential for nucleon projectile, while for the multi-nucleon projectiles is the weighted sum of proton and neutron basis-generating WS potentials  \cite{mercenne_2019,michel_book_2021}:
\begin{eqnarray}
\!\!\!\!\!\!\!\!\!\!\!\! U^{ L }_{{\rm CM}, {\rm C}} (r) &=& z~U^{ L }_{\rm p, C} (r) + n~U^{ L }_{\rm n, C} (r)  
\label{UCM_central} \\
\!\!\!\!\!\!\!\!\!\!\!\! U^{ L }_{\rm CM, SO } (r) &=& \frac{z}{n+z}~U^{ L }_{\rm p, SO} (r) + \frac{n}{n+z}~U^{ L }_{\rm n, SO} (r) 
 \label{UCM_so_average} \ ,
\end{eqnarray}
where $z$ and $n$ are the number of protons and neutrons of the cluster, respectively, while $U^{ L }_{\rm p, C} (r)$, $U^{ L }_{\rm p, SO} (r)$ and $U^{ L }_{\rm n, C} (r)$, $U^{ L }_{\rm n, SO} (r)$ are the WS basis-generating central and spin-orbit potentials for protons and neutrons, respectively. 
One uses fractional masses in front of neutron and proton spin-orbit potentials in order to form an average spin-orbit potential in Eq.~(\ref{UCM_so_average}). Indeed, from this averaging procedure, one can separate the radial spin-orbit dependence from the sum over the 
$\mathbf{\ell} \cdot \mathbf{s}$ terms of the nucleons of the cluster projectile. The spin of the projectile then appears explicitly : $$\sum \mathbf{\ell} \cdot \mathbf{s} \simeq (1/(n+z)) ~ \mathbf{L}_{\rm CM}  \cdot \sum \mathbf{s} \simeq (\mathbf{L}_{\rm CM} \cdot \mathbf{J_{\rm int}})/(n+z) \ .$$ 
An additional factor $1/(n+z)$ arises because $\ell \simeq L_{\rm CM}/(n+z)$ \cite{mercenne_2019,michel_book_2021}.
The potential $U^{ L }_{\rm CM}(r)$ of Eq.~(\ref{HCM_definition}) then reads:
\begin{eqnarray}
U^{ L }_{\rm CM} (r) &=& U^{ L }_{\rm CM, C} (r) \nonumber \\
&+& \frac{1}{n+z} ~ { U^{ L }_{\rm CM, SO }} (r) ~ (\mathbf{L}_{\rm CM} \cdot \mathbf{J_{\rm int}}) \ .   
\label{UCM}
\end{eqnarray}

In order to calculate the kernels $H_{\rm cc' } (r,r')$ and $N_{ \rm cc' } (r,r')$ (Eqs.~(\ref{h_cc_compound}) and (\ref{n_cc_compound})), one expands ${ | (c,r) } >$ onto a one-body Berggren basis:
\begin{equation}
  | ({\rm c},r) > = \sum_{N_{\rm CM}} \frac{ { u }_{ N_{\rm CM}} (r) }{ r } | ({\rm c},N_{\rm CM}) > \ ,
  \label{expansion_channel_n}
\end{equation}
where $N_{\rm CM}$ refers to the projectile c.m.~shell number in the Berggren basis generated by diagonalizing ${ { \hat{ H } }_{\rm CM } }$ (see Eqs.~(\ref{HCM_definition}) and (\ref{UCM})), i.e.~${ { \hat{ H } }_{\rm CM } | N_{\rm CM} ~ L_{\rm CM} > = { E }_{\rm  CM } | N_{\rm CM} ~ L_{\rm CM} } >$,
and where $${ | ({\rm c},N_{\rm CM}) > = \hat{ \mathcal{A}} | \{ |\Psi_{ \rm T }^ {J_{ \rm T } }> \otimes |N_{\rm CM} ~ L_{\rm CM} ~ J_{\text{int}} ~ J_{\rm P} \}_{ M }^{J}} > \ .$$ For simplicity, the channel dependence has been omitted in the notation of $u_{N_{\rm CM}}(r)$.

Consequently, using Eq.~(\ref{expansion_channel_n}) one can expand Eqs.~(\ref{h_cc_compound}) and (\ref{n_cc_compound}) onto the basis of ${ | ({\rm c},N_{\rm CM}) } >$  and derive the Hamiltonian and norm kernels:
\begin{eqnarray}
  H_{\rm cc' } (r, r') &=& \left( { \hat{ H } }_{\rm CM } + E_{\rm T }^{ { J }_{\rm T } } + {E}_{\rm P}^{ { J }_{ \text{int} } } \right) \frac{ \delta (r - r') }{ r r' } { \delta }_{\rm cc' } \nonumber \\
  &+&  { \tilde{ V }}_{\rm cc' } (r , r')
  \label{hamiltonian_matrix_elmts} \\
  N_{\rm cc' } (r,r') &=&  \frac{ \delta (r - r') }{ r r' } { \delta }_{\rm cc' }  + \Delta N_{\rm cc' } (r,r') \ ,
  \label{norm_matrix_elmts}
\end{eqnarray}
where $\tilde{ V }_{\rm cc' }(r,r')$ which includes the remaining short-range potential terms of the Hamiltonian kernels is the inter-cluster potential, and $\Delta N_{\rm cc' } (r,r')$ is a finite-range operator as well. $H_{\rm cc' } (r, r')$ reduces to its diagonal part at large distances as $\tilde{ V }_{\rm cc' }(r,r')$ vanishes identically at a large but finite radius outside the target. Hence, the nucleon transfer which is induced by $\tilde{ V }_{\rm cc' }(r,r')$, and consequently ${ { \hat{ H } }_{\rm TP } }$, acts only occur in the vicinity of the target and not in the asymptotic region.

The determination of ${ { \tilde{ V } }_{\rm cc' }(r,r') }$ involves the calculation of the matrix elements of ${ { \hat{ H } }_{\rm TP } }$.
In order to compute ${ { \hat{ H } }_{\rm TP } }$ one has to expand each ${ | ({\rm c},N_{\rm CM}) } >$ onto a basis of Slater determinants built upon single-particle (s.p.) states of the Berggren ensemble.
In practice, the intrinsic target ${ | { \Psi }_{\rm T }^{ { J }_{\rm T } } } >$ and projectile ${ | { J }_{\rm int } } >$ states are already calculated with that basis, as ${ { \hat{ H } }_{\rm T } }$ and ${ { \hat{ H } }_{ \text{int} } }$ are solved using the GSM. 
Consequently, the nuclear structure of target and projectile states is more realistic than in the R-matrix theory, for example, where nuclei are typically built from a few uncorrelated clusters \cite{Descouvemont_2010}.
Note that, as we deal with light projectiles, ${ { \hat{ H } }_{ \text{int} } }$ is solved within a no-core shell model framework.
The remaining task consists in expanding ${ | N_{\rm CM} ~ L_{\rm CM} ~ { J }_{\rm int } } >$ in a basis of Slater determinants. In GSM-CC, one applies for this a c.m. excitation raising operator onto ${ | { J }_{\rm int } } >$. More details can be found in Refs.\cite{michel_book_2021,Linares_2023}.

The many-body matrix elements of the norm kernel Eq.~(\ref{n_cc_compound}) are calculated using the Slater determinant expansion of the cluster wave functions ${ | ({\rm c},N_{\rm CM}) } >$.
The antisymmetry of channels which is enforced by the antisymmetrizer in Eq.~(\ref{channel}), is exactly taken into account through the expansion of many-body targets and projectiles with Slater determinants.
The treatment of the non-orthogonality of channels is the same as in the one-nucleon projectile case \cite{jaganathen_2014}. Indeed, 
the channels are not orthogonal in general because of the antisymmetrizer in Eq.~(\ref{channel}). In order to generate orthogonal channels, the norm kernel
$N_{\rm cc' } (r , r')$ (see Eq.~(\ref{n_cc_compound})) has to be diagonalized. The orthogonalized channels are linear combinations of the initial channels $| \left( {\rm c} , r \right)>$ (see Eq.~\ref{channel}). The channel equations for orthogonalized channels then become:
\begin{eqnarray}
  \sum_{\rm c}\int_{0}^{\infty}  \!\!\! && r^{ 2 } \left( \widetilde{H}_{\rm cc' } (r , r') - E \delta_{cc'} \frac{\delta(r-r')}{rr'} \right) \nonumber \\
  &\times& \frac{ { w }_{\rm c } (r) }{ r } = 0	\ ,
  \label{cc_cluster_eq_orthog}
\end{eqnarray}
where $\tilde{H}_{\rm cc' } (r , r')$ contains $H_{\rm cc' } (r , r')$ and terms induced by the orthogonalization of channels, and the ${ w }_{\rm c } (r)$ functions are the orthogonalized channel wave functions \cite{jaganathen_2014}.
Once the kernels are computed, the coupled-channel equations of Eq.~(\ref{cc_cluster_eq}) can be solved using a numerical method based on a Berggren basis expansion of the Green's function ${ { (H - E) }^{ -1 } }$, that takes advantage of GSM complex energies. Details of this method can be found in Refs. \cite{mercenne_2019,michel_book_2021}.

\subsection{Wigner cusps in reaction cross-sections and spectroscopic factors}
Wigner derived the threshold law for cross-sections \cite{wigner1948behavior} by analyzing the asymptotic behavior of the wave functions.    Later, discussion of the Wigner cusps was done in the $R$-matrix theory. \cite{breit1957energy,baz1958energy,newton1958inelastic,fonda1961inelastic,meyerhof1963threshold,baz1969scattering,lane1970theory}. 
For the endoergic reactions (e.g. production of slow neutral particles) one finds:
\begin{equation}
            \label{eq:abovethresh}
            \sigma (i\rightarrow j) \sim k_j^{2\ell_j+1} \sim E_j^{\ell_j + 1/2 }\, .
      \end{equation}
        For exoergic reactions (e.g. absorption of slow neutrons) one obtains:
        \begin{equation}
            \label{eq:belowthresh}
            \sigma (i\rightarrow j) \sim k_i^{2\ell_i-1} \sim E_i^{\ell_i - 1/2 }\, .
        \end{equation}
       From Eq. (\ref{eq:abovethresh}) and (\ref{eq:belowthresh}), one can see that there is a discontinuity (cusp) for $\ell=0,1$. 
    For higher $\ell$ values, the cross-section is smooth and the discontinuity shows up in the $(\ell-1)$th derivative of the cross-section with respect to the energy of the system.  

 For charged particles the dependence of the cross-section near threshold is more complicated:
  For endoergic reactions it is:
        \begin{equation}
            \label{eq:abovethreshproton}
            \sigma(i\rightarrow j) \sim e^{-2\pi\eta_j} \ ,
        \end{equation}
        whereas for an exoergic reactions one finds:
       \begin{equation}
            \label{eq:belowthreshproton}
            \sigma(i\rightarrow j) \sim k^2_ie^{-2\pi\eta_i} \ ,
        \end{equation}
    where $\eta$ is the Sommerfeld parameter. Eqs. (\ref{eq:abovethreshproton}) and (\ref{eq:belowthreshproton}) provide useful relations to extrapolate cross-sections for charged particles to low bombarding energies. 
    
    Due to the unitarity of the S-matrix, the opening of a new reaction channel removes some of the flux from other reaction channels, hence modifying the cross-section in other channels. 
     
  Wigner cusps are not related to any particular physical system or interaction among its constituents. They have been found experimentally  also in carbon nanotubes \cite{blue2022observation}, in scattering of positrons by He atoms \cite{caradonna2012search}, in electron photodetachment or in collisions of ultracold atoms \cite{sadeghpour2000collisions}, in the $\pi^- p \rightarrow\pi^0n$ reaction at the opening of the $\eta$ channel \cite{starostin2005measurement}, and in many other systems (see Refs. \cite{batley2006observation,adair1958production}).   
  
 Wigner cusps in the SFs are the consequences of unitarity which is rigorously respected in the GSM and GSM-CC approaches. At the particle emission threshold, some of the probability flux goes to the new channel(s) that in turn changes the occupancy of single-particle shells  \cite{michel2007continuum}.  Different dependencies on energy distance from the decay threshold imply that the mirror many-body systems at the respective neutron and proton dripline will, in general, have different SFs.

\section{From threshold effects to clusterization}  

    Structure of the near-threshold states and hence the universality of the clustering phenomenon cannot be the consequence of a specific feature of the nuclear interaction because then the nucleon-nucleon correlations or clustering correlations in near-threshold states would appear at random.  The experimental data are telling us the contrary, by providing countless examples of the close relation between the structure of a near-threshold state and the nature of a nearby channel. In the OQS approach, all shell model states with the same quantum numbers couple to the same given decay channel.     This induces a mixing among them, which may lead to the formation of the collective eigenstate, the so-called \textit{aligned eigenstate}, which couples strongly to the decay channel and carries an imprint of its features.
  An indication for the enhanced probability of near-threshold states comes also from the $R$-matrix analysis which points out that the increased density of levels with large reduced width close to the particle emission threshold favors the formation of the near-threshold resonance  \cite{barker1964model}. Moreover, the close relation between the level density and the phase shifts:
   \begin{equation}
       \rho_\ell (E) = \frac{1}{\pi} \frac{d \delta_\ell (E)}{dE} \ ,
    \end{equation}
suggests connection between Wigner cusps in the cross-sections and the appearance of the near-threshold states, as one would expect a sharp change in phase shifts close to the resonance or the antibound state. 

 The key question is: how to characterize the configuration mixing in OQSs? One way is to study avoided crossings of resonances that are associated with the exceptional points in the resonance spectrum~\cite{zirnbauer1983destruction,okolowicz2009}. However, 
  a more convenient way is to calculate the suitably defined correlation energy~\cite{Okolowicz2003,Okolowicz2012,Okolowicz2013}.  
  In shell model embedded in the continuum, one defines a continuum-coupling correlation energy as the difference between the effective Hamiltonian $\mathcal{H}_{QQ}$, which includes effects of the coupling to the scattering continuum on discrete shell model states, and the closed quantum system (CQS) shell model Hamiltonian $H_{\rm SM} \equiv H_{QQ}$, in a certain OQS eigenvector $|\Psi_i>$ of the $\mathcal{H}_{QQ}$:
    \begin{equation}
   	\label{blabla}
        E_{{\rm corr,}i}^{(\ell)} (E) = <\Psi_i|\mathcal{H}_{QQ}(E) - H_{QQ}|\Psi_i> \ .
    \end{equation}
    Here, $E_{{\rm corr,}i}^{(\ell)}$ is the correlation energy in the OQS eigenstate $i$ with angular momentum $\ell$ \cite{Okolowicz2003,Okolowicz2012,Okolowicz2013}.  As seen in the above expression, the idea is that the continuum-coupling correlation energy  $E_{{\rm corr,}i}^{(\ell)} (E)$ captures the essential features of the coupling between discrete and continuum states and thus provides an account of the interplay between the continuum-coupling induced configuration mixing and the effects of the Coulomb and centrifugal potentials. 
   Indeed, it was found that the maximum of the correlation energy is at the threshold for $\ell=0$ neutrons. For protons, the effect is shifted above the threshold due to the Coulomb barrier \cite{Okolowicz2012}. Minimum of the correlation energy $E_{{\rm corr,}i}^{(\ell)} (E)$ indicates the optimal energy where the coupling to the decay channel is strongest \cite{Okolowicz2012,Okolowicz2013} and the probability appearance of the collective state aligned with the decay channel is most probable.
In GSM-CC, one can define the continuum-coupling correlation energy in a given many-body eigenstate $|\Psi^{J^{\pi}}>$ as a difference of energies:
  \begin{equation}
        \label{eq:corrEforGSM}
        E_{{\rm corr,}c}^{J^{\pi}} = <\Psi^{J^{\pi}}|H|\Psi^{J^{\pi}}> - <\Psi^{J^{\pi}}_{ \cancel{c}} |H|\Psi^{J^{\pi}}_{ \cancel{c} }> = E_{\rm full}^{J^{\pi}} - E_{ \cancel{c} }^{J^{\pi}} \ ,
    \end{equation}
    where $|\Psi^{J^{\pi}}>$ is the full GSM-CC solution defined as an expansion in terms of all reaction channels, and $ |\Psi^{J^{\pi}}_{ \cancel{c}}> $ is the wave function expansion without the reaction channel $c$. $E_{\rm full}^{J^{\pi}}$ and $E_{ \cancel{c} }^{J^{\pi}}$ in Eq. (\ref{eq:corrEforGSM}) are the energy eigenvalues corresponding to the eigenfunctions $|\Psi^{J^{\pi}}>$ and $ |\Psi^{J^{\pi}}_{ \cancel{c}}>$, respectively.

\subsection{Model space and Hamiltonian}
\label{model_space}
{
The lowest particle emission thresholds in $^7$Be and $^7$Li are $^4$He + $^3$He and  $^4$He + $^3$H, respectively. Hence, both $^3$He ($^3$H): 
$[{^4}{\rm He}\otimes{^3}{\rm He}]^{J^{\pi}}$ ($[{^4}{\rm He}\otimes{^3}{\rm H}]^{J^{\pi}}$),
 and proton (neutron): {$[{^6}{\rm Li}((J^\pi)_T)\otimes{\rm p}(\ell j)]^{J^{\pi}}$ ($[{^6}{\rm Li}((J^\pi)_T)\otimes{\rm n}(\ell j)]^{J^{\pi}})$}
 channels will be used to describe the low-energy states of $^6$Li, $^7$Be ($^7$Li). In the considered reaction channels for  $^7$Be and $^7$Li, $^4$He is always in its ground state $0^+$, so  the angular momentum and parity of $^3$He and $^3$H clusters are equal to $J^{\pi}$.  Conversely, the total angular momentum and parity $J_{\rm T}$ and $\pi_{\rm T}$ of $^6$Li depend on the structure of $^6$Li, as $^6$Li can be in its ground or excited states.  }
 
   \begin{figure*}[htb] 
                \begin{center}
                \includegraphics[width=12cm,angle=0]{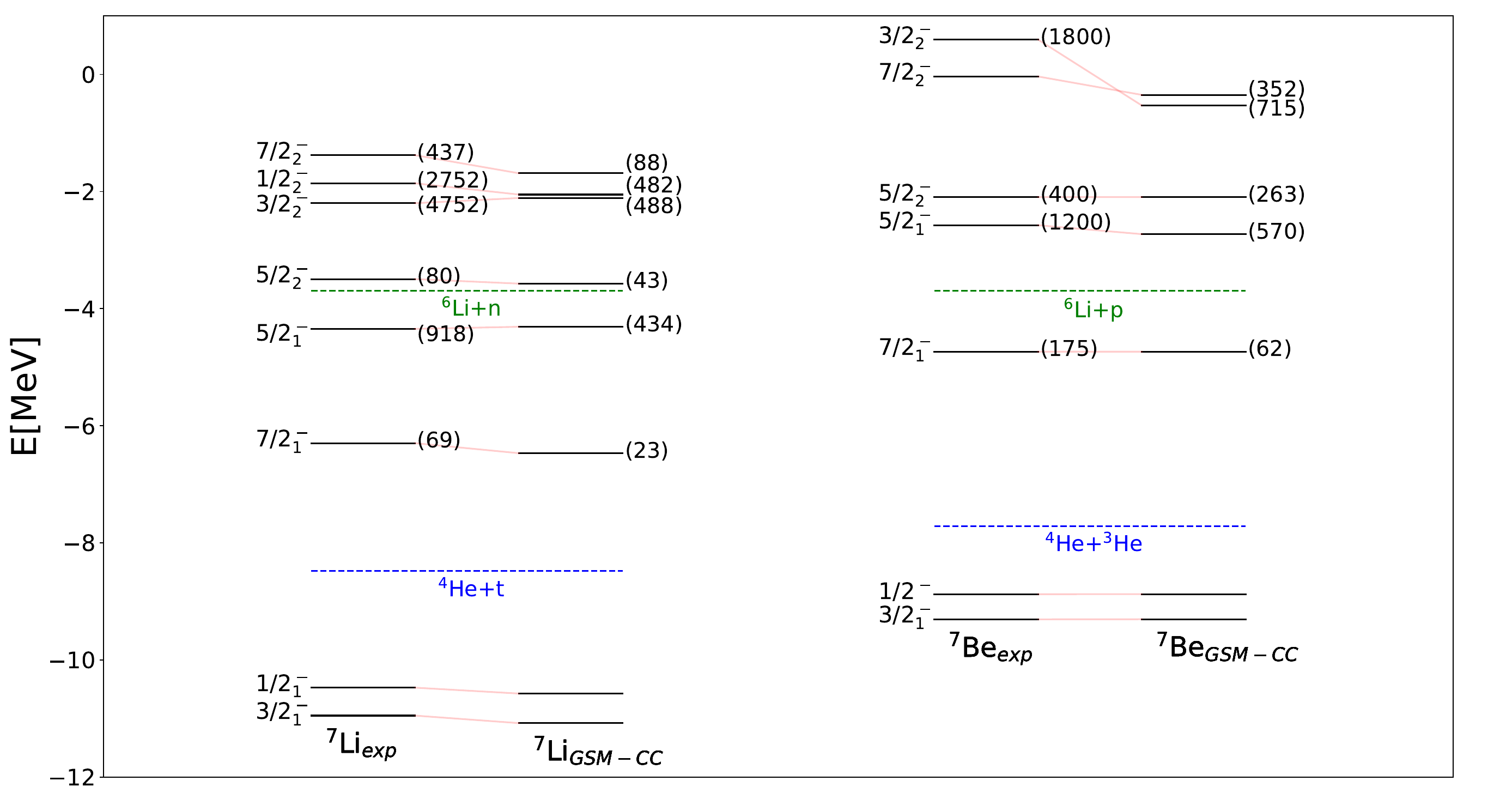}
                \end{center}
                %\vskip -1.5truecm
                \caption{(Color online) The calculated energy spectra  of $^{7}$Li and $^7$Be, are compared with the experimental data \cite{spiger1967scattering}. Numbers in the brackets indicate the resonance width in keV.}
                \label{spec_7Li7Be}
            \end{figure*} 
The effective Hamiltonian is optimized using the GSM in Berggren basis. The model is formulated in the relative variables of COSM that allow to eliminate spurious c.m.~excitations (see Sect.\ref{Formalism}).
We use $^4$He as an inert core with two, three or four valence nucleons to describe $^6$Li, $^7$Be and $^7$Li wave functions, respectively. 
The Hamiltonian consists of the one-body part and the nucleon-nucleon interaction of FHT type~\cite{1979Furutani} supplemented by the Coulomb term: %$V_{\rm FHT} = V_{\rm c} + V_{\rm LS} + V_{\rm T} + V_{\rm Coul} $~ ,
\begin{equation}
V_{\rm FHT} = V_{\rm c} + V_{\rm LS} + V_{\rm T} + V_{\rm Coul}  \ ,
\label{V_FHT}
\end{equation}
where $V_{\rm c}$ , $V_{\rm LS}$, $V_{\rm T}$ represent its central, spin-orbit and tensor part, respectively. The two-body Coulomb potential 
$V_{\rm Coul}(r) = e^2/r$ between valence protons is treated exactly at GSM-CC level by incorporating its long-range part into the basis potential. 
(see Ref.\cite{PRC_isospin_mixing} for a detailed description of the method). 
 
The $^4$He core is mimicked by a one-body potential of the WS type, with a spin-orbit term, and a Coulomb field. 
The WS potential depth $V_0$, the spin-orbit strength $V_{\ell s}$, the radius $R_0$, and the diffuseness $a$ are the four parameters that enter the optimization carried out independently for protons and neutrons. The Coulomb potential is kept fixed and equal to the potential generated by a spherical Gaussian charge distribution: $U_{\rm Coul}(r)=2e^2{\rm erf}(r/{\tilde R}_{\rm ch})/r$~\cite{Sai77},
where ${\tilde R}_{\rm ch} = 4R_{\rm ch} /(3\sqrt{\pi} )$ and ${\rm erf}(r/{\tilde R}_{\rm ch})$ is the error function in $r/{\tilde R}_{\rm ch}$. The previous value for ${\tilde R}_{\rm ch}$ allows $U_{\rm Coul}(r)$ to resemble the Coulomb potential generated by a uniformly charged distribution of radius $R_{\rm ch}$. 

The internal structure of $^3$He and $^3$H projectiles in the channels $[{^4}{\rm He}(0^+_1)\otimes{^3}{\rm He}(L_{\rm CM}~J_{\rm int}~J_{\rm P})]^{J^{\pi}}$ and $[{^4}{\rm He}(0^+_1)\otimes{^3}{\rm H}(L_{\rm CM}~J_{\rm int}~J_{\rm P})]^{J^{\pi}}$ is calculated using the N$^3$LO interaction~\cite{PRC_N3LO} without the three-body contribution, fitted on phase shifts properties of proton-neutron elastic scattering reactions. 
The N$^3$LO realistic interaction is diagonalized in six HO shells to generate the intrinsic states of $^3$He  and $^3$H. The oscillator length in this calculation is $b=1.65$ fm.  In the coupled-channel equations of the GSM-CC, we use the experimental binding energies of $^3$He, $^3$H to assure correct thresholds $^4$He + $^3$He, $^4$He + $^3$H.
The use of two different interactions to deal with the structure of $^7$Be ($^7$Li) and the elastic scattering of $^3$He ($^3$H) on $\alpha$-particle is necessary as we have two different pictures in our model. Before and after the reaction, $^3$He ($^3$H) is far from the target and its properties as a cluster projectile are prominent, whereas during the reaction the properties of composite systems $^7$Be ($^7$Li) are decisive. As the FHT interaction is defined from $^6$Li, $^7$Be, $^7$Li properties, it cannot grasp the structure of $^3$He ($^3$H) at large distances.
Conversely, the N$^3$LO interaction cannot be used in a core and valence particles approximation. 
As the N$^3$LO interaction enters only the $^3$He ($^3$H) projectile basis construction, it is not explicitly present in the Hamiltonian, but just insures that the projectile $^3$He ($^3$H) has both the correct wave function (binding energy) and asymptotic behavior. 
Details concerning the effective Hamiltonian and the model space in GSM-CC calculation can be found in Ref. \cite{Linares_2023}. 

The relative motion of the $^3$He ($^3$H) cluster c.m.~and the $^4$He target is calculated in the Berggren basis generated  by proton and neutron Woods-Saxon potentials. 
Different $L_{\rm CM}=0,1,2,3$ partial waves bear 3, 3, 2, 2 pole states, respectively, which are included along with the respective contours. 
    
 Figure \ref{spec_7Li7Be} shows the GSM-CC spectrum of $^{7}$Li and $^7$Be. All energies of the states are given relative to the energy of $^4$He core.  One may notice a change in the order of higher lying levels $1/2^-_2$, $3/2^-_2$, $7/2^-_2$ between $^7$Be and $^7$Li due to different threshold energies and Coulomb energies.

\section{Near threshold effects and the continuum-coupling correlation energy}
\label{Results}

 In this section we will present results of the GSM-CC in the multi-mass partition formulation for the orthogonal reaction channel probabilities and the continuum-coupling correlation energy for selected states of $^{7}$Be and $^7$Li.  An example is shown in figure \ref{7Li5|2-_2} for the $5/2^-_2$ state in $^7$Li. Here the orthogonal channel probability weights for the states $|{ \Psi }^{ 5/2^-_2 } >$, $|{ \Psi }^{ 5/2^-_2 }_{\cancel{c}} >$ and the correlation energy $E_{\rm corr}^{5/2^{-}_2}$, are shown as a function of the energy distance from the lowest one-neutron decay threshold $[{^6}{\rm Li}(1^+_1)\otimes{\rm n}(\ell j)]^{5/2^{-}}$.  We vary the distance of a state to the threshold by changing the Woods-Saxon depth $V_0$. 

\begin{figure}
\centering
\sidecaption
\includegraphics[width=6cm,clip]{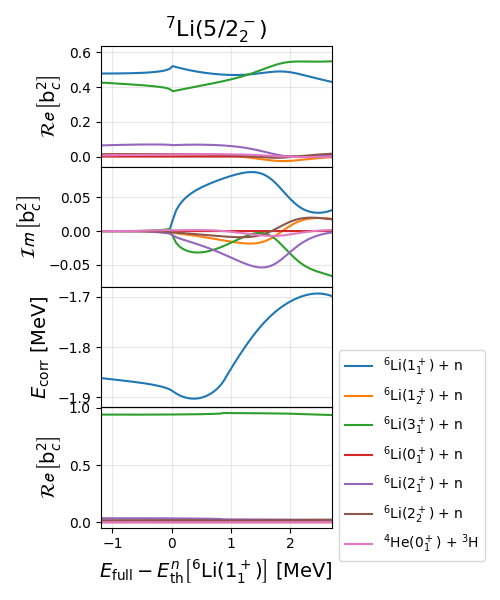}
\caption{The GSM-CC analysis for the state $|{ \Psi }^{ 5/2^-_2 } >$ in $^7$Li. From top to bottom: (i) the real part of the channel weights ${\cal R}e[b_c^2]$, (ii) the imaginary part of the channel weights ${\cal I}m[b_c^2]$, (iii) the correlation energy $E_{\rm corr}^{5/2^{-}_2}$, and (iv) the real part of the channel weights ${\cal R}e[b_c^2]$ for the state $|{ \Psi }^{ 5/2^-_2 }_{\cancel{c}} >$, are shown as a function of the distance with respect to the one-neutron decay threshold $[{^6}{\rm Li}(1^+_1)\otimes{\rm n}(\ell j)]^{5/2^{-}}$.}
\label{7Li5|2-_2}       % Give a unique label
\end{figure}
\begin{figure}
\centering
\sidecaption
\includegraphics[width=6cm,clip]{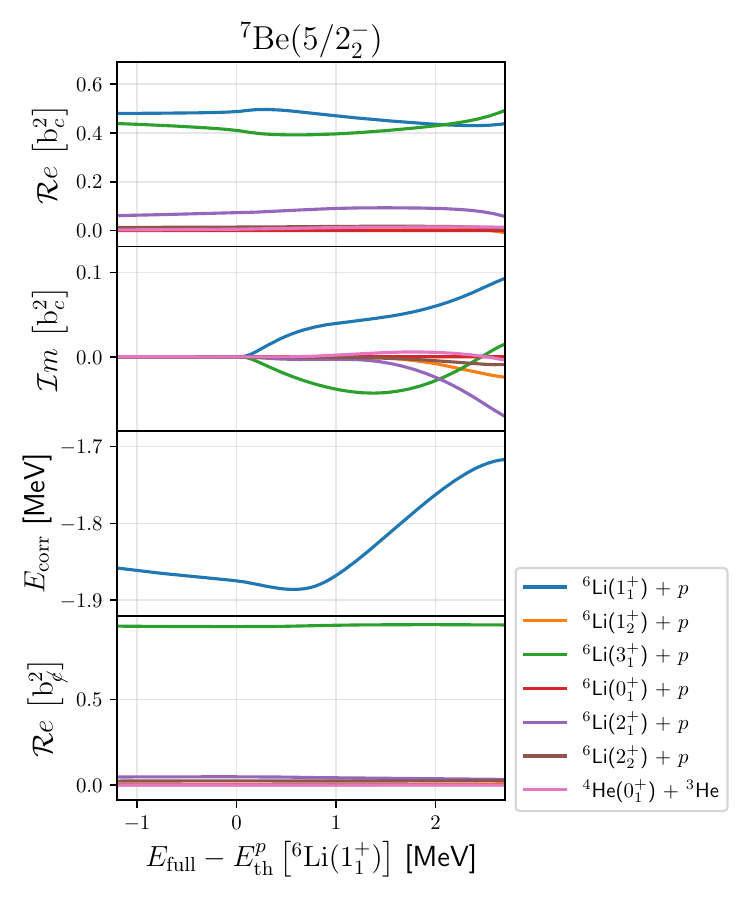}
\caption{The same as in figure \ref{7Li5|2-_2} but in a mirror nucleus $^7$Be. All quantities are plotted as a function of the distance from the one-proton decay threshold $[{^6}{\rm Be}(1^+_1)\otimes{\rm p}(\ell j)]^{5/2^{-}}$.}
\label{7Be5|2-_2}       % Give a unique label
\end{figure}

	 The orthogonal channel probability weights are defined by:
            \begin{equation}
            	<\Psi|\Psi>  = \sum\limits_{c} \left| <w_c|w_c> \right|^2 = \sum\limits_{c} b_c^2 = 1 + 0i \, .
            \end{equation}
            The sum over all channels of the real parts of channel weights ${{\cal R}e}[b_c^2]$ is normalized to 1, whereas the sum of imaginary parts ${{\cal I}m}[b_c^2]$ is equal to 0.   In the complex-energy framework of the GSM-CC, the squared wave-function amplitude (channel weight) ${\cal R}e [b_c^2]$ can be negative \cite{michel_review_2009}.   At the leading order, the statistical uncertainty of the ${\cal R}e [b_c^2]$ is associated with its imaginary part ${\cal I}m [b_c^2]$ \cite{michel_book_2021}}.  ${\cal R}e [b_c^2]$ is the average value of the corresponding channel probability in different measurements, while ${\cal I}m [b_c^2]$ can be related to the dispersion rate over time in the measurement, and hence represents its statistical uncertainty \cite{michel_book_2021}.    	

              In the state $|{ \Psi }^{ 5/2^-_2 }_{\cancel{c}} >$, the channel $[{^6}{\rm Li}(1^+_1)\otimes{p(\ell j)}]^{5/2^-}$ is absent.
        	Probability of the neutron channel $[{^6}{\rm Li}(1^+_1)\otimes{\rm n}(\ell j)]^{5/2^{-}}$ in the state $|{ \Psi }^{ 5/2^-_2 } >$ grows while approaching the channel threshold and exhibits a cusp slightly above the neutron threshold. 
        	The minimum of the correlation energy associated with this channel is shifted by about 0.35 MeV above the threshold. 
        	This shift between the maximum of the real part of the channel weight and the minimum of the corresponding correlation energy which is seen in figure \ref{7Li5|2-_2}, might be related to a different behavior of the imaginary part of the channel weight which has a maximum shifted to higher energies with respect to the maximum of the real part.
        	
\begin{figure}
\centering
\sidecaption
\includegraphics[width=6cm,clip]{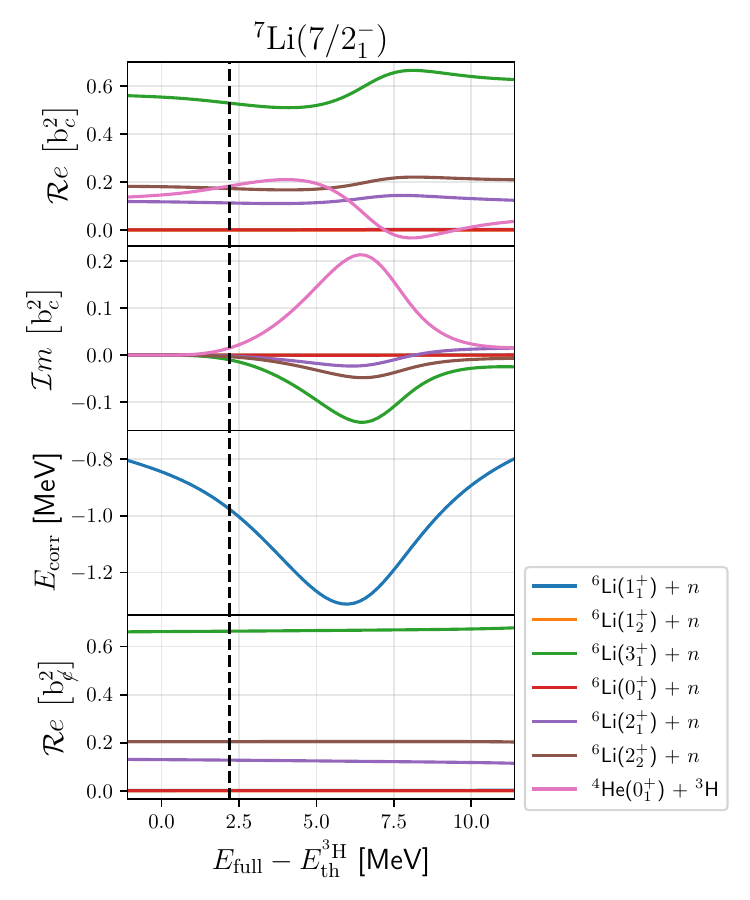}
\caption{The GSM-CC analysis for the state $|{ \Psi }^{ 7/2^-_1 } >$ in $^7$Li. From top to bottom: (i) the real part of the channel weights ${\cal R}e[b_c^2]$, (ii) the imaginary part of the channel weights ${\cal I}m[b_c^2]$, (iii) the correlation energy $E_{\rm corr}^{7/2^{-}_1}$, and (iv) the real part of the channel weights ${\cal R}e[b_c^2]$ for the state $|{ \Psi }^{ 7/2^-_1 }_{\cancel{c}} >$, are shown as a function of the distance with respect to the triton decay threshold $^4$He + $^3$H. The dashed vertical line gives the GSM-CC energy for the $7/2^-_1$ resonance.}
\label{7Li7|2^-_1_clus}       % Give a unique label
\end{figure}

\begin{figure}
\centering
\sidecaption
\includegraphics[width=6cm,clip]{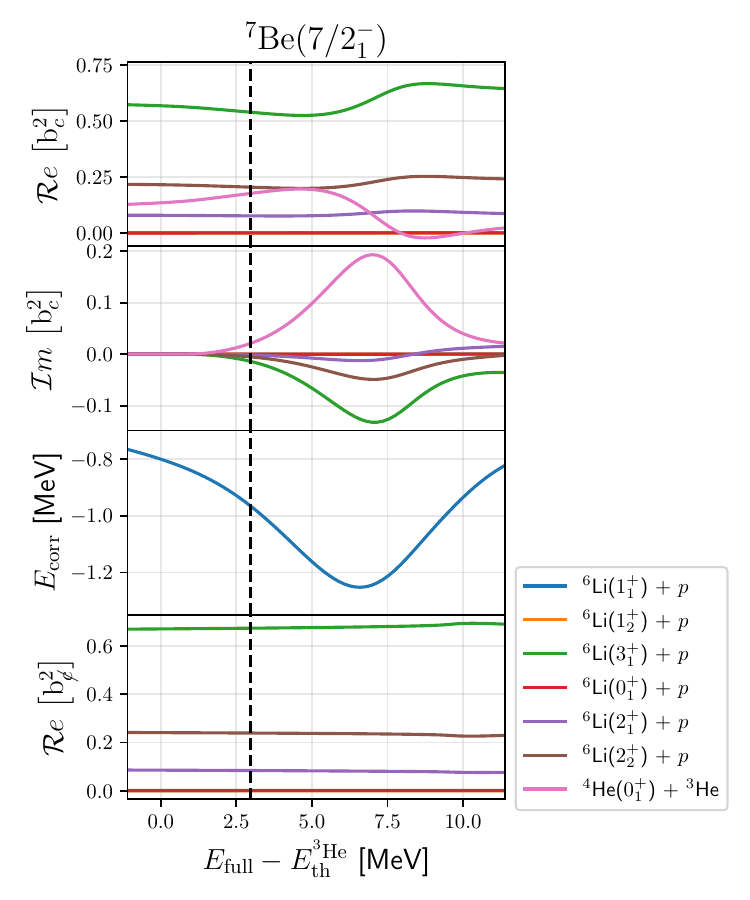}
\caption{The same as in figure \ref{7Li7|2^-_1_clus} but for a $7/2^-_1$ state in the mirror nucleus $^7$Be. All quantities are plotted as a function of the distance with respect to the $^3$He decay threshold $^4$He + $^3$He. The dashed vertical line gives the GSM-CC energy for the $7/2^-_1$ resonance.}
\label{7Be7|2^-_1_clus}       % Give a unique label
\end{figure}

        	Figure \ref{7Be5|2-_2} shows the GSM-CC results for $5/2^-_2$ state of $^7$Be, a mirror of the state $5/2^-_2$ in $^7$Li. One can see that the mirror symmetry is well satisfied.  The position of minima of the correlation energy in mirror states differ by about $\sim$0.2 MeV. 
        	This is consistent with the observation done in the $7/2^-_1$ states (see Figs. \ref{7Li7|2^-_1_clus} and \ref{7Be7|2^-_1_clus}).
        	
            Figure \ref{7Li7|2^-_1_clus} presents dependence on the energy difference with respect to the lowest decay threshold 
            $^4$He + $^3$H of the real and imaginary parts of the channel weights $b_c^2$ in the state $|{ \Psi }^{ 7/2^-_1 } >$ of $^7$Li, the real part of the channel weights $b_c^2$ in the state $|{ \Psi }^{ 7/2^-_1 }_{\cancel{c}} >$ calculated without the triton channel $[{^4}{\rm He}(0^+_1)\otimes{^3{\rm H}}(L_{\rm CM}~J_{\rm int}~J_{\rm P})]^{7/2^-}$, and the correlation energy $E_{\rm corr}^{7/2^{-}_1}$ (see Eq. (\ref{eq:corrEforGSM})). As discussed previously, the state $7/2^-_1$ has a pronounced $^3$H-cluster structure. 
            In the whole interval of energies, neutron reaction channel $[{^6}{\rm Li}(3^+_1)\otimes{\rm n}(\ell j)]^{7/2^{\pi}}$ provides a smaller contribution to the wave function $7/2^-_1$. 
            Probability of the $^3$H channel $[{^4}{\rm He}(0^+_1)\otimes{^3{\rm H}}({{}^{2}F_{7/2}})]^{7/2^-}$ in $|{ \Psi }^{ 7/2^-_1 } >$ grows when approaching the emission threshold and has a maximum $\sim$4.5 MeV above the threshold energy $E_{\rm th}^{{^{3}{\rm H}}}$. 
            The minimum of the correlation energy is seen at $\sim 6$ MeV above the cluster emission threshold, which is intermediate between the maxima of ${\cal R}e[b_c^2]$ and ${\cal I}m[b_c^2]$ for the channel $^4$He + $^3$H. 
            Removal of the triton channel $[{^4}{\rm He}(0^+_1)\otimes{^3{\rm H}}(^2F_{7/2})]^{7/2^-}$ in $|{ \Psi }^{ 7/2^-_1 }_{\cancel{c}} >$ leaves the real part of the wave function nearly unchanged, bearing a small increase towards $\sim 9$ MeV where we approach the proton emission threshold. However, this effect is small and should not interfere with the minimum of the correlation energy.
            
            Results for the mirror state are shown in figure \ref{7Be7|2^-_1_clus}. This figure shows a dependence of the real and imaginary parts of the channel weights $b_c^2$ in the state $|{ \Psi }^{ 7/2^-_1 } >$, the real part of the channel weights $b_c^2$ in the state $|{ \Psi }^{ 7/2^-_1 }_{\cancel{c}} >$ without the $^3$He channel $[{^4}{\rm He}(0^+_1)\otimes{^3{\rm He}}(^2F_{7/2})]^{7/2^-}$, and the correlation energy $E_{\rm corr}^{7/2^{-}_1}$, on the energy difference with respect to the $^3$He decay threshold 
            $^4$He + $^3$He in $^7$Be. 
            One can see that the mirror symmetry in $7/2^-_1$ states of $^7$Li and $^7$Be is obeyed and the dependence on energy difference from the respective thresholds in $^7$Li and $^7$Be is qualitatively the same. 
            There is in fact a small deviation of $\sim0.3$ MeV of the minima of $^7$Li and $^7$Be. 
            This can be attributed to a stronger Coulomb interaction in the channel $[{^4}{\rm He}(0^+_1)\otimes{^3{\rm He}}({{}^{2}F_{7/2}})]^{7/2^-}$.

\section{Discussion}
\label{discussion}
{

Nuclear states form a network of coupled states communicating with each other through decays and captures~\cite{Okolowicz2003,dobaczewski2007}. If the continuum states are neglected, this communication is broken and each system becomes an isolated closed quantum system (CQS). It is obvious, that the CQS description of atomic nuclei (e.g. the nuclear shell model) becomes self-contradictory for weakly-bound or unbound states.
A classic example of a continuum coupling is the Thomas-Ehrman shift which manifests itself in the asymmetry in the energy spectra between mirror nuclei having different particle emission thresholds. 
A consistent description of the interplay between scattering states, resonances, and bound state requires an OQS formulation.
The proximity of the branching point at threshold induces a collective mixing in shell-model-like states mediated by the aligned state that shares many features of the decay channel. 
This salient phenomenon in open quantum systems has been studied for the first time using the GSM-CC~\cite{Linares_2023}. 
The signature of a profound change of the near-threshold shell model wave function and of the direct manifestation of the continuum-coupling induced correlations is the presence of cluster states near their corresponding cluster emission thresholds~\cite{Linares_2023}. The appearance of clustering in a given reaction channel  is associated with the collective response in other channels due to unitarity. 
In addition to the special effects of cluster correlations, one also expects a significant modification of in-band and intra-band  electromagnetic transitions in the proximity of the open particle emission threshold~\cite{Ploszajczak2020}.

We have shown here on examples of few eigenstates in $^7$Li, $^7$Be how $^3$H, $^3$He cluster correlations appear when the shell model-like state approaches cluster decay threshold, and how these correlations faint again further away from threshold.
We have applied the multi-mass-partition GSM-CC approach. The lowest threshold in these nuclei corresponds to the emission of clusters of nucleons and therefore the GSM-CC description of resonance wave functions requires the inclusion of the reaction channels involving both clusters and nucleons. With the two mass partitions $^6$Li+n, $^4$He+$^3$H for $^7$Li, $^6$Li+p, and $^4$He+$^3$He for $^7$Be, we obtain a good description of resonance energies and widths. 
Extensive studies in shell model embedded in the continuum~\cite{Okolowicz2012,Okolowicz2013} have demonstrated that the low-energy coexistence of the cluster-like and shell-model-like configurations can be reconciled in the open quantum system formulation of shell model. 
The convenient measure of the continuum-coupling induced near-threshold collectivization is the continuum-coupling correlation energy, as defined in Eq. (\ref{eq:corrEforGSM}). 
In the studies using shell model embedded in the continuum, it was demonstrated that the continuum-coupling energy correction in general rises with the number of valence nucleons and is bigger in even-even isotopes~\cite{okolowicz2007continuum}. 
We have shown in GSM-CC calculation a significant increase of this energy correction in the vicinity of the open reaction channel. 
The excess of the collective energy due to the coupling to a nearby decay channel varies depending on the nature of the threshold and the structure of the close-lying shell model eigenstate which is dressed by correlations due to the  mixing with other shell model states of the same quantum numbers. In the studied cases of $^7$Li, $^7$Be, the surplus of collective continuum-coupling energy in a near-threshold OQS eigenstate varies from $\sim$200 keV to $\sim$600 keV and is larger for  the coupling to cluster decay channels: $^4$He+$^3$H for $^7$Li, $^6$Li+p, and $^4$He+$^3$He for $^7$Be. 

Amount of this  collective energy may seem at first insignificant but an example of nuclear pairing shows that even if the correlation energy is small as compared to the nuclear binding energy, nevertheless it might have crucial influence on the structure of nucleus and on the nucleon-nucleon correlations. This is also the case of continuum-coupling correlations which at each opening of the particle emission threshold show significant correlation energy variations.
The spectacular effects of the continuum-coupling correlation energy can be seen in the vicinity of any particle emission threshold even though the effect is predicted to be strongest at low excitation energies. 
The typical energy scale and the strength of the coupling varies depending on the angular momentum involved in the decay channel and the strength of the Coulomb interaction. It should be remembered that the total continuum-coupling energy is significantly larger than the individual contribution from the coupling to the nearby decay channel. In the studied case of $^7$Li, $^7$Be OQS eigenstates, this energy varies from $\sim$1.2 MeV for $5/2^-_2$ state and the neutron/proton decay channels to $\sim$1.9 MeV for $7/2^-_1$ state and the $^3$H/$^3$He decay channels.

Near-threshold phenomena are \textit{terra incognita} of the nuclear physics. Systematic experimental and theoretical studies of the (i) collectivization of wave functions due to the coupling to decay channel(s), (ii) formation of clusters/correlations: $^2$H, $^3$H, $^3$He, 3n, 4n, ..., which carry an imprint of nearby decay channel(s), (iii) modification of the spectroscopic factors and its consequence for the nucleosynthesis, (iv) coalescing resonances and their effects in nuclear structure and reactions, etc. are necessary in the future. In all of those phenomena, unitarity plays an essential role. 
The unitarity is the fundamental property of Quantum Mechanics, yet the mainstream nuclear theory works in the unitarity violating schemes. 
With the advent of the shell model for OQS, the quantitative studies of near-threshold phenomena in nuclear structure and reactions become now possible.

\vskip 0.5truecm
\section{Acknowledgments}
Discussions with Witek Nazarewicz are gratefully acknowledged. 
One of us (J.P.L.F.) wish to thank the GANIL laboratory for a hospitality when most of the results of this paper have been obtained. This work has been supported by the National Natural Science Foundation of China under Grant Nos. 11975282; the Strategic Priority Research Program of Chinese Academy of Sciences under Grant No. XDB34000000; the State Key Laboratory of Nuclear Physics and Technology, Peking University under Grant No. NPT2020KFY13. 

 \bibliography{refs.bib}

\end{document}